\documentclass[sigconf]{acmart}
\usepackage{graphicx}
\AtBeginDocument{%
  \providecommand\BibTeX{{%
    \normalfont B\kern-0.5em{\scshape i\kern-0.25em b}\kern-0.8em\TeX}}}

\setcopyright{acmcopyright}
\copyrightyear{2021}
\acmYear{2021}
\acmDOI{10.1145/1122445.1122456}

\acmConference[SIGIR '21]{SIGIR '21: The 44th International
ACM SIGIR Conference on Research and Development in Information Retrieval}{July 11--15, 2021}{Montreal, Quebec, Canada}
\acmBooktitle{SIGIR '21: The 44th International ACM SIGIR Conference on Research and Development in Information Retrieval,
  July 11--15, 2021, Montreal, Quebec, Canada}
\acmPrice{15.00}
\acmISBN{978-1-4503-XXXX-X/18/06}

\begin{document}

\title{Distillation based Multi-task Learning: A Candidate Generation Model for Improving Reading Duration}

\author{Zhong Zhao, Yanmei Fu, Hanming Liang, Li Ma, Guangyao Zhao, Hongwei Jiang}
\affiliation{
  \institution{Tencent Inc.}
  \city{Shenzhen}
  \country{China}
}
\email{{zhongzhao,friedafu,meeloliang,listma,issaczhao,rockyjiang}@tencent.com}

\renewcommand{\shortauthors}{Zhao and Fu, et al.}

\begin{abstract}
In feeds recommendation, the first step is candidate generation. Most of the candidate generation models are based on CTR estimation, which do not consider user's satisfaction with the clicked item. Items with low quality but attractive title (i.e., click baits) may be recommended to the user, which worsens the user experience. One solution to this problem is to model the click and the reading duration simultaneously under the multi-task learning (MTL) framework. There are two challenges in the modeling. The first one is how to deal with the zero duration of the negative samples, which does not necessarily indicate dislikes. The second one is how to perform multi-task learning in the candidate generation model with double tower structure that can only model one single task. In this paper, we propose an distillation based multi-task learning (DMTL) approach to tackle these two challenges. We model duration by considering its dependency of click in the MTL, and then transfer the knowledge learned from the MTL teacher model to the student candidate generation model by distillation. Experiments conducted on dataset gathered from traffic logs of Tencent Kandian's recommender system show that the proposed approach outperforms the competitors significantly in modeling duration, which demonstrates the effectiveness of the proposed candidate generation model.
\end{abstract}


%\begin{CCSXML}
%<ccs2012>
%<concept>
%<concept_id>10002951.10003317.10003338.10003343</concept_id>
%<concept_desc>Information systems~Learning to rank</concept_desc>
%<concept_significance>500</concept_significance>
%</concept>
%</ccs2012>
%\end{CCSXML}




\maketitle

\section{Introduction}
Click through rate (CTR) estimation is a widely adopted method for ranking in many recommender systems. Many models based on deep learning are proposed to estimate CTR in recent years\cite{WDL,DeepFM,DCN,DIN}. As for feeds recommendation, the predicted CTR (pCTR) can reflect how likely the user will click the item, but can not reflect how likely the user will like the item after click and read the content. For example, item with low quality and attractive titles (i.e. click baits) usually gets a high pCTR but users never like them. Therefore, only modeling CTR can not ensure the user's satisfaction with the clicked item. To improve user's experience, reading duration should be modeled as well, which is of great importance in industrial applications such as feeds recommendation.

In this paper, we focus on reading duration modeling and its application to large-scaled candidate generation for feeds recommendation. There are two main challenges in our real practice. The first challenge is how to deal with the zero duration of the negative samples. The negative samples get zero duration just because they are un-clicked, which doesn't necessarily indicate that the user dislikes the item. It is quite different from the positive samples with short duration which indicates dislike. Directly use the zero duration as target for modeling may lead to the inaccurate estimation. The second challenge is caused by the first challenge. In order to solve the problem of the first challenge, multi-task learning is employed. However, it is difficult to perform multi-task learning in the candidate generation model. As we know, most deep learning based candidate generation models adopt the double tower structure\cite{DSSM,Youtube}. They have a user tower and a item tower for computing the user vector and item vector respectively, and use user-item inner product as similarity metric for ANN search to generate candidate items in a very efficient way. Since the inner product can only model one single task, it makes multi-task learning difficult to be applied directly to the candidate generation model. To our knowledge, few papers discuss duration modeling. In the real practice, the commonly used method is to model duration by regression in one single task, in which the duration of all negative samples are set to zero and square loss is used. As mentioned before, fitting the duration of negative sample to zero may treat the dislikes (short duration) and the un-clicks (zero duration, but not necessarily dislike) similarly, which may mislead the model training.

To tackle the challenges stated above, we propose a \textbf{d}istillation based \textbf{m}ulti-\textbf{t}ask \textbf{l}earning approach, which we refer to as \textbf{DMTL}, to model reading duration for candidate generation. We overcome the problems of the existing duration models by considering duration's dependency to click within the multi-task learning framework which simultaneously models CTR and CTCVR for the click task and the duration task respectively. Then, we use the distillation technique to transfer knowledge learned by the multi-task model to the double tower candidate generation model, which makes the candidate generation model obtain the ability of modeling reading duration while keeping its high efficiency in candidate generation.

To evaluate the performance of the proposed approach, we conducted experiments on the dataset gathered from the traffic logs of Tencent Kandian's recommender system. The results of the offline and online experiments show that the proposed approach outperforms the competing duration models significantly, which demonstrates the effectiveness of the proposed approach in modeling reading duration for candidate generation.

\begin{figure*}[htbp]
\centering
\includegraphics[width=0.85\textwidth]{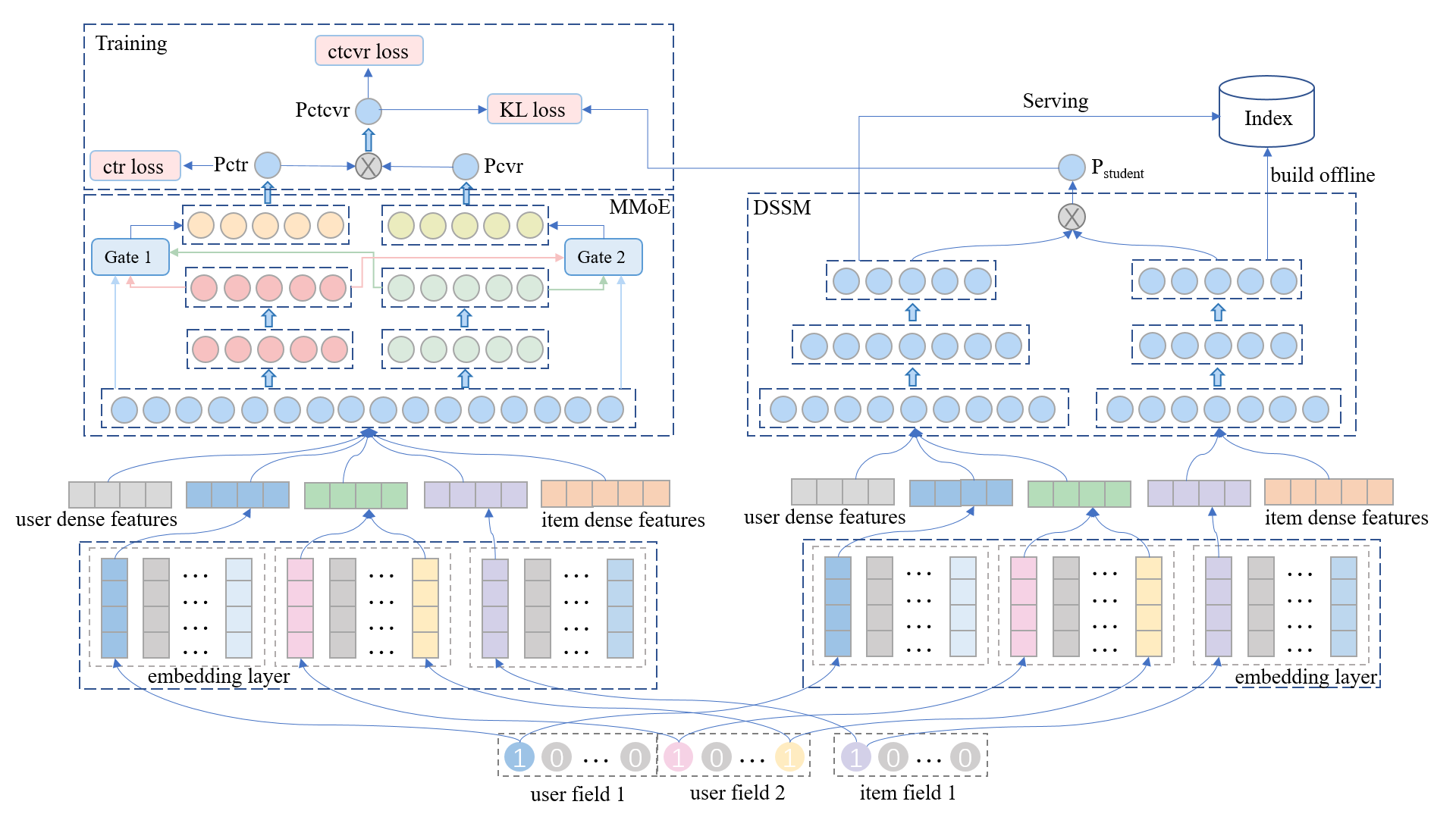}
\caption{Network structure of the proposed distillation based multi-task learning model. The teacher model (left) is a multi-task learning model which models reading duration. It considers the dependency of click and reading by minimizing the CTR loss and CTCVR loss simultaneously. The student model (right) is a candidate generation model with double tower structure. Knowledge of the teacher model is transferred to the student model by distillation so that the student model can obtain the ability to model reading duration while keeping its high efficiency for candidate generation.}
\label{fig_network_structure}
\end{figure*}

\section{The proposed approach}
\subsection{Multi-task Learning for Duration Modeling}
\label{teacher_model}
The purpose of candidate generation is to select hundreds or thousands of items that are relavant to user's interests from the whole item corpus which may have millions or even billions of items. In this paper, the proposed DMTL improves the quality of candidate generation by modeling click and reading duration simultaneously, rather than modeling click only. For the click task, positive samples are clicked impressions, and negative samples are randomly selected from all items according to their frequency of being clicked. This is quite different from ranking model which uses clicked impressions as positive sample and un-clicked impressions as negative samples. For the duration task, positive samples are clicked impressions with duration more than 50 seconds (i.e., the median of all durations), and the rest are negative samples.

Let $u_i$ and $v_i$ be the user features and item features respectively, both of which are usually concatenation of embeddings of multiple fields, and $x_i$ be the concatenation of $u_i$, $v_i$ and other dense features. Let $y_i$ be the label of click task with $y_i=1$ representing the item is clicked and $y_i=0$ representing the item is randomly selected. Let $z_i$ be the label of duration task with $z_i=1$ representing the item has been read for more than 50s, and $z_i=0$ otherwise. The duration modeling problem can be formulated as estimating the probability of $z_i=1$ given $x_i$, i.e., $p(z_i=1|x_i)$. As mentioned before, $z_i$ is dependent of $y_i$ since $y_i=0$ will cause $z_i$=0. To better model this probability, we make good use of the dependency of click and reading. Specifically, $p(z_i=1|x_i)$ can be rewritten as
\begin{equation}
\label{CTCVR_Factorization}
p(z_i=1|x_i) = p(y_i=1|x_i)p(z_i=1|y_i=1, x_i)
\end{equation}
where $p(y_i=1|x_i)$ is the predicted click-through rate(pCTR), $p(z_i=1|y_i=1, x_i)$ is the predicted conversion rate(pCVR) and $p(z_i=1|x_i)$ is the predicted click-through and conversion rate (pCTCVR). To reduce the influence of selection bias and data sparsity when modeling duration, we adopt the approach proposed in ESMM\cite{ESMM} which fits CTR and CTCVR simultaneously under the multi-task learning framework. In our model, click task and duration task fit the CTR and the CTCVR respectively by minimizing the binary cross entropy.

We employ the multi-task learning framework MMoE\cite{MMoE,YoutubeMTL} to model CTR and CVR. Let $f_k$ be the $k$-th expert network which is usually a DNN and $f_k(x_i)$ be the output vector of the $k$-th expert. For the modeling of CTR, the gate is computed as $g_c(x)=[g_{c1}(x_i), \cdots, g_{cK}(x_i)]$, where $g_c(\cdot)$ is the gate function defined as $g_c(x)=softmax(W_cx_i)$ with $W_c$ being a trainable matrix, $K$ is the number of experts, and $g_{ck}(x_i)$ is the $k$-th element of $g_c(x_i)$. The output of the experts for modeling CTR is computed as
\begin{equation}
\label{MOEClick}
e_c(x_i) = \sum_{k=1}^Kg_{ck}(x_i)f_k(x_i)
\end{equation}
For the modeling of CVR, the gate function $g_d(\cdot)$ with trainable parameter matrix $W_d$ can be defined similarly, and the output of the experts for modeling CVR is computed as
\begin{equation}
\label{MOEDuration}
e_d(x_i) = \sum_{k=1}^Kg_{dk}(x_i)f_k(x_i)
\end{equation}
where $g_{dk}$ is the $k$-th element of $g_d(x_i)$.
The pCTR and pCVR for sample $x_i$ are modeled as
\begin{equation}
\label{pctr}
p_{ctr}(x_i, \theta_t) = sigmoid(h_c(e_c(x_i)))
\end{equation}
\begin{equation}
\label{pcvr}
p_{cvr}(x_i, \theta_t) = sigmoid(h_d(e_d(x_i)))
\end{equation}
where $h_c(\cdot)$ and $h_d(\cdot)$ are DNNs that map $e_c(x_i)$ and $e_d(x_i)$ to the logit of pCTR and pCVR respectively, and $\theta_t$ is all trainable parameters in the above formulations. According to \ref{CTCVR_Factorization}, \ref{pctr} and \ref{pcvr}, the pCTCVR can be written as
\begin{equation}
\label{pctcvr}
p_{ctcvr}(x_i, \theta_t) = p_{ctr}(x_i, \theta_t)p_{cvr}(x_i, \theta_t)
\end{equation}
As modeling reading duration is to fit the CTCVR, the loss of the duration task is the following binary cross entropy
\begin{equation}
\label{loss_duration}
L_d(\theta_t) = -\sum_{i=1}^Nz_i\log p_{ctcvr}(x_i, \theta_t) + (1-z_i)\log(1-p_{ctcvr}(x_i, \theta_t))
\end{equation}
Equation \eqref{pctcvr} and \eqref{loss_duration} have modeled the dependency of click and reading by introducing $p_{ctr}(x_i, \theta_t)$ to compute $p_{ctcvr}(x_i, \theta_t)$. However, only fitting $p_{ctcvr}(x_i, \theta_t)$ to CTCVR can not ensure $p_{ctr}(x_i, \theta_t)$ fits to CTR. Therefore, we need a auxiliary task to make sure that $p_{ctr}(x_i, \theta_t)$ fits to the CTR. The loss function of this auxiliary click task is formulated as
\begin{equation}
\label{loss_click}
L_c(\theta_t) = -\sum_{i=1}^Nz_i\log p_{ctr}(x_i, \theta_t) + (1-z_i)\log(1-p_{ctr}(x_i, \theta_t))
\end{equation}
By summing \eqref{loss_duration} and \eqref{loss_click}, we get the multi-task learning loss function for the duration model as
\begin{equation}
\label{loss_teacher}
L_{teacher}(\theta_t) = w_1 L_d(\theta_t) + w_2 L_c(\theta_t)
\end{equation}
where $w_1$ and $w_2$ are the weights for each loss.

\subsection{Distillation for Candidate Generation}
In most deep learning based candidate generation models, double tower structure is employed to compute user vectors and item vectors, where the item vectors are used to build the item index. For a given user vector, the user-item inner product is used as similarity for ANN search in the item index, and the top-k items are returned as candidate items. However, the candidate generation models are unable to model duration by multi-task learning since the inner product can only model one task. To make the candidate generation model obtain the extra ability of modeling reading duration within its high efficient double tower structure framework, we use distillation technique to transfer knowledge learned by the MTL model in section~\ref{teacher_model} to the candidate generation model.

The proposed candidate generation model uses the double tower structure, and computes the user vector and item vector by DNNs. Let $R(u_i)$ and $S(v_i)$ be the user vector and the item vector respectively, where $R(\cdot)$ and $S(\cdot)$ are the DNNs that map input embedding to output vector. Given $R(u_i)$ and $S(v_i)$, the CTCVR predicted by the candidate generation model can be formulated as
\begin{equation}
\label{student_ctcvr}
p(z_i=1|R(u_i), S(v_i), \theta_s) = sigmoid(R(u_i)^TS(v_i))
\end{equation}
where $R(u_i)^TS(v_i)$ is the inner product of $R(u_i)$ and $S(v_i)$, and $\theta_s$ is the trainable parameters in $R(u_i)$ and $S(v_i)$. We expect that $p(z_i=1|R(u_i), S(v_i), \theta_s)$ can be similar to $p_{ctcvr}(x_i, \theta_t)$ as much as possible, so that we can use \eqref{student_ctcvr} to estimate the CTCVR of duration accurately while keeping the high efficiency of the candidate generation model. To this end, we treat the multi-task learning model \eqref{loss_teacher} as the teacher model and the double tower candidate generation model \eqref{student_ctcvr} as the student model, and use distillation to transfer knowledge from the teacher model to the student model. The loss of the distillation can be formulated as the following KL-divergence
\begin{equation}
\begin{split}
\label{loss_kl}
L_{student}(\theta_s) = p_{ctcvr}(x_i, \theta_t)\frac{p_{ctcvr}(x_i, \theta_t)}{p(z_i=1|R(u_i), S(v_i), \theta_s)} \\
+ (1-p_{ctcvr}(x_i, \theta_t))\frac{1-p_{ctcvr}(x_i, \theta_t)}{1-p(z_i=1|R(u_i), S(v_i), \theta_s)}
\end{split}
\end{equation}
By summing the loss of the teacher model and the student model, we obtain the loss of the distillation-based multi-task learning model as follow
\begin{equation}
\label{loss_total}
L(\theta_t, \theta_s) = L_{teacher}(\theta_t) + L_{student}(\theta_s)
\end{equation}
To prevent the teacher model from being influenced by the student model in the training stage, the parameters in the student model are separated from those in the teacher model, and the pCTCVR of teacher model is freezed when computing $L_{student}(\theta_s)$. Therefore, minimizing loss \eqref{loss_total} is equivalent to minimizing the teacher loss and the student loss alternatively. In the inferring stage, we only use the student model to compute user vectors and item vectors, where item vectors are used for building the index, user vector is used as the query, and ANN search is performed to fetch top-k candidate items from the index for the user. The network structure and trainging/serving framework of the proposed model is illustrated in figure \ref{fig_network_structure}.

\section{Experiment}
\subsection{Offline experiment}
\subsubsection{Dataset}
Experiments are conducted on dataset gathered from traffic logs of Tencent Kandian's feeds recommender system. The dataset has billions of training samples and millions of test samples. For each user, the positive samples are the clicked impressions, and the negative samples are randomly selected from all items according to their frequency of being clicked. Each clicked impression has a positive reading duration, and each randomly selected item has a zero duration.

\subsubsection{Competitors}
We compare the proposed DMTL to the existing candidate generation models. The competitors are listed as follow

\textbf{DSSM-Regression}: user vectors and item vectors are computed by DNNs, and their inner product regresses the reading duration by square loss. The duration of the negative samples is treated as zero for training.

\textbf{DSSM-Classification}: user vectors and item vectors are computed by DNNs, and their inner product is used to compute binary cross entropy for training a classification model, where positive samples are clicked-impressions with reading duration more than 50s and the rest are negative samples.

\textbf{DSSM-Click}: user vectors and item vectors are computed by DNNs, and their inner product is used to compute binary cross entropy for training a classification model, where positive samples are clicked-impressions and the rest are the negative samples.

\subsubsection{Metric}
We compare all approaches by evaluating their performance in a binary classification task, in which positive samples are true clicked-impressions with duration larger than 50s and the rest are negative samples. Area under ROC curve (AUC) is used as metric for performance evaluation. Higher AUC represents stronger ability of modeling reading duration.

\subsubsection{Parameter settings}
For the teacher model, the experts are DNNs with hidden layer size $1024\times512\times256$ and the number of experts is 2. The DNN for each task is $256\times256$. For the student model, the hidden layer size for each tower is $512\times256\times128$. For both models, the embedding size for the categorical variables is 30.

\subsubsection{Results and Analysis}
Table \ref{table_offline_experiment} shows the performance of each candidate generation model. The regression model performs the worst among all models, which may be due to its direct fitting to large number of zero durations. The classification model is a little better than the regression model but still performs much poor than the proposed DMTL. This is due to the lack of modeling dependency between click and reading, which leads to the confusion between un-clicks and short duration. Compared to the duration DSSM, the click DSSM performs much better, which indicates the importance of click to duration. As the occurrence of duration depends on click, modeling duration without considering the dependency of click may miss a lot of important information for training. Among all models, the proposed DMTL achieves the highest AUC and significantly outperforms the competing methods. The improvement is attributed to the knowledge distilled from the teacher model which models duration in a more reasonable way that considers dependency of the click.

\begin{table}[!ht]
\centering
\caption{Offline performance of different duration models.}
\begin{tabular}{lc}
\toprule
models & AUC\\
\midrule
DSSM-Regression & 0.7374\\
DSSM-Classification & 0.7562\\
DSSM-Click & 0.9359\\
\textbf{DMTL} & 0.9544\\
\bottomrule
\end{tabular}
\label{table_offline_experiment}
\end{table}

\subsection{Online experiment}
Online A/B tests are also conducted to compare the proposed DMTL to the competing candidate generation models. For the online experiment, we only change the candidate generation step by using different candidate generation models and keep all other steps unchanged. The online evaluation metric is the average reading duration defined as $T/M$, where $T$ is the total sum of all reading durations (seconds) and $M$ is total number of impressions. Table \ref{table_online_experiment} shows the online performance of different models. DSSM-Regression and DSSM-Classification perform much worse than the DSSM-click and DMTL, which is consistent with the result of the offline experiment. Directly modeling duration without modeling its preceding step (click) may lead to the inaccurate estimation which causes the returned candidate items to be less relavant to user's interests. The proposed DMTL overcomes this problem, and thus achieves the best performance compared to competing methods.

\begin{table}[!ht]
\centering
\caption{Online performance of different duration models.}
\begin{tabular}{lc}
\toprule
models & average reading duration (s)\\
\midrule
DSSM-Regression & 2.73\\
DSSM-Classification & 1.74\\
DSSM-Click & 3.93\\
\textbf{DMTL} & 4.27\\
\bottomrule
\end{tabular}
\label{table_online_experiment}
\end{table}

\section{Conclusion and Future Work}
In this paper, we proposed a distillation based multi-task learning approach for modeling reading duration in candidate generation stage. The teacher model is a multi-task learning model which models reading duration by using ESMM to consider the dependency of click and reading. The student model is a DSSM candidate generation model with double tower structure. Knowledge Distillation technique is employed to make DSSM obtain the ability of modeling reading duration while keeping its high efficiency in candidate generation. Offline/online experiments conducted on real world dataset demonstrated the effectiveness of the proposed approach in modeling reading duration for candidate generation. The proposed approach can be easily generalized to other scenario in which there are multiple tasks with dependency. In the future, we will study how to develop multi-task learning model in the case when only part of the tasks are related, and how to fuse the output score for distillation.

\bibliographystyle{ACM-Reference-Format}
\bibliography{references}

\end{document}